\documentclass{iaus}
\usepackage{graphicx}

\def\simgr{\,\hbox{\hbox{$ > $}\kern -0.8em \lower 1.0ex\hbox{$\sim$}}\,}
\def\simle{\,\hbox{\hbox{$ < $}\kern -0.8em \lower 1.0ex\hbox{$\sim$}}\,}

\title[Evolution before the ZAMS] 
{Stellar evolution before the ZAMS}

\author[F. Palla]   
{Francesco Palla$^1$}%

\affiliation{$^1$INAF-Osservatorio Astrofisico di Arcetri, Largo E. Fermi 5,
50125, Firenze, Italy\break email: palla@arcetri.astro.it}

\pubyear{2005}
\volume{227}  
\pagerange{119--126}
\date{}
\jname{Massive Star Birth: A Crossroads of Astrophysics}
\editors{R. Cesaroni, E. Churchwell, M. Felli \& M. Walmsley, eds.}
\begin{document}

\maketitle

\begin{abstract}
Young stars on their way to the ZAMS evolve in significantly different ways
depending on their mass. While the theoretical and observational properties
of low- and intermediate-mass stars are rather well understood and/or
empirically tested, the situation for massive stars (\simgr 10--15~M$_\odot$)
is, to say the least, still elusive. On theoretical grounds, the PMS
evolution of these objects should be extremely short, or nonexistent at all.
Observationally, despite a great deal of effort, the simple (or bold)
predictions of simplified models of massive star formation/evolution have
proved more difficult to be checked. After a brief review of the theoretical
expectations, I will highlight some critical test on young stars of various
masses.
\keywords{stars: formation - Hertzsprung-Russell diagram - stars: pre--main-sequence}
\end{abstract}

\firstsection 
\section{Introduction}

After a brief, intense, and optically obscured protostellar phase, newly
formed stars emerge from their natal environments as relatively bright,
distended objects undergoing gravitational contraction towards the main
sequence.  While residual accretion from circumstellar disks is observed to
take place in young pre--main-sequence (PMS) stars, the magnitude of the
accretion flow appears to be much lower (typically, $\dot M_{\rm
PMS}\simle 10^{-7}$~M$_\odot$~yr$^{-1}$) than the protostellar one
(typically, $\dot M_{\rm proto}\simle 10^{-5}$~M$_\odot$~yr$^{-1}$).
Therefore, PMS derive most of their energy from contraction rather than
accretion, and evolve on a characteristic Kelvin-Helmoltz time scale ($t_{\rm
KH}$).  The major imprint left from the protostellar phase on PMS stars is
their specific initial conditions.  At the end of the main accretion phase,
the stellar core first appears as an optically visible star along the
so-called birthline (\cite{sta83}).  Empirically, the birthline provides the
initial conditions for PMS contraction, specifying the internal structure and
radius of the newly formed star.  Depending on the mass of the star, young
visible stars are classified as T Tauri (low-mass, typically $\simle$1-2
M$_\odot$), or Herbig Ae/Be stars (intermediate-mass, $\simle$10 M$_\odot$).
For $\dot M_{\rm proto}\simle 10^{-5}$~M$_\odot$~yr$^{-1}$, the birthline
joins the ZAMS at M$_\ast\sim$8--10 M$_\odot$ (\cite{ps90}).  Thus, for stars
with masses above this value there should be no optical PMS phase: the rapid
gravitational contraction of massive protostellar cores has caused hydrogen
ignition already during the accretion phase. Therefore, massive stars are
expected to be born {\it on} the ZAMS.

This brief excursus highlights the physical link between
protostellar and PMS evolution and justifies the notion that young massive
stars have a distinctively different early evolution than the lower mass
counterparts. To date, there is still a lack of numerical models capable of
following these two phases self-consistently.  While
the powerful simulations described in these proceedings (see the
contributions by Bonnell and Klessen) can follow the fragmentation,
collapse and accretion flows of dense gas within cluster-like
conditions, they cannot resolve the innermost regions where the protostellar
core grows in mass and eventually turns into a PMS object.  Thus, in order to
describe the main features of stellar evolution before the ZAMS, we have
still to rely on calculations that either ignore completely the protostellar
phase as in the classical models of PMS evolution (e.g., 
\cite{bar98}, \cite{sie00}) or inherit the initial conditions from
independent simplified treatments of the accretion flow. In the following, I
will summarize the main properties of the latter class of models, referring
to \cite{pal02} for a full discussion. Then, I will discuss separately some
important tests of the predictions of PMS models covering the 
full mass spectrum.

\section{The HR diagram and PMS evolution}

A key property of protostellar evolution is the existence of a {\it
mass-radius} relation for accreting cores. This relation finds its physical
origin in the thermostatic nature of deuterium burning once the protostar
reaches the critical ignition temperature  ($\sim 10^6$~K).  As a
result of deuterium burning, first at the center and then in a subsurface
shell, protostars never attain large radii: the radius remains typically 3--5
times larger than the corresponding value on the ZAMS, but a factor of ten,
or more, smaller than the large values predicted by the classical theory of
PMS evolution (Iben, Hayashi, Cameron, etc.) that ignored the impact of
protostellar evolution. 

Another important feature of protostars is that their internal structure
departs significantly from the assumption of thermal convection. In the
standard PMS theory, all stars were assumed to be fully convective objects as
a result of the large initial radii. However, in low-mass protostars
convection is due to central deuterium burning, while protostars more massive
than about 2 M$_\odot$ are radiatively stable in the inner regions, and
possess a thick convective mantle maintained by deuterium burning.  Thus,
these stars are {\it thermally unrelaxed} at the beginning of the PMS
evolution and must undergo {\it non-homologous} quasi-static contraction.

\begin{figure}
 \includegraphics[height=12cm,width=8cm,angle=-90]{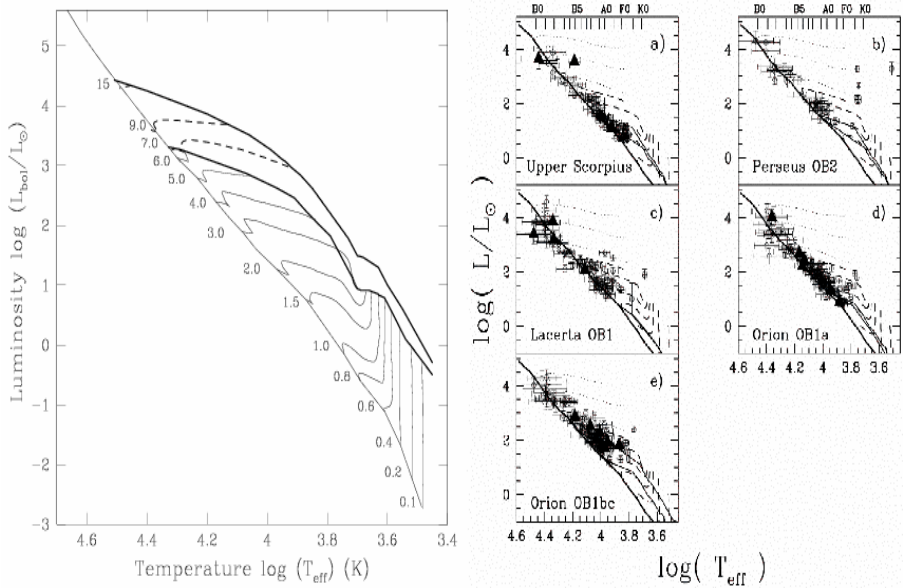}
  \caption{{\it Left panel}: Location of the stellar birthline in the HR 
  diagram for accretion rates
  of $10^{-5}$ (lower heavy curve) and $10^{-4}$ (upper heavy curve)
  M$_\odot $~yr$^{-1}$. Labeled curves (both solid and dashed) are PMS 
  evolutionary tracks for the indicated stellar mass.
  {\it Rigth panel}: Distribution of A- and B-type stars in several nearby
  OB associations. (From \cite{her05})}
  \label{hrd}
\end{figure}

These two properties have a profound impact on the calculation of PMS
evolutionary models and the resulting HR diagram. First, compared with the
classical set of PMS tracks, those shown in Fig.~\ref{hrd} occupy a much more
limited portion of the diagram. The reason is that the older tracks
correspond to stellar radii too large to be attained during protostellar
accretion. Thus, the starting luminosities (proportional to $R_\ast^2$) are
much lower than previously envisioned. In addition, the surface temperature
of these stars begin {\it higher}. 
Notice also that the birthline intersects the main sequence
at a mass M$_\ast\sim8$~M$_\odot$: this point represents the critical stellar
configuration in which hydrogen burning has stopped gravitational contraction
while the star is still growing in mass.  {\it More massive stars, therefore,
have no PMS phase at all, but appear directly on the main sequence once they
are optically revealed}. The exact mass at which the birthline merges into
the ZAMS depends on the accretion history and increases as $\dot M_{\rm acc}
\uparrow$, as shown in Fig.~\ref{hrd}. 

\subsection{Empirical tests on the HR diagram}

\begin{figure}
\centering
 \includegraphics[height=7cm,width=7cm]{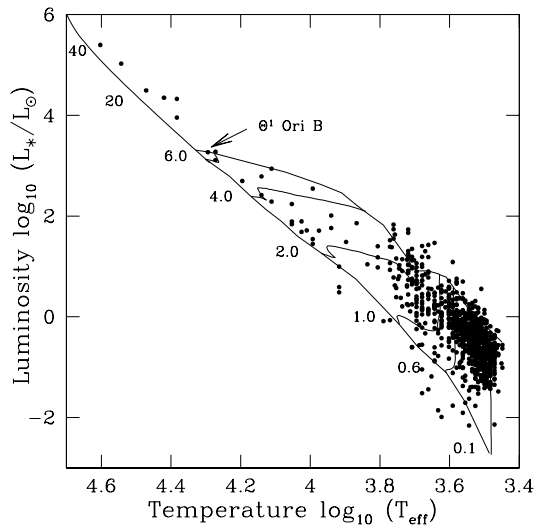}
  \caption{HR diagram of the ONC (from~\cite{hil97}). The position of 
  the eclipsing binary $\Theta^1$~Ori~B star (discussed in Sect. 5) 
  is indicated.}
  \label{onchrd}
\end{figure}

An important consequence is that observed PMS stars should
not be located to the right of the birthline. This prediction can be readily
tested by the observations of large samples of T Tauri and Herbig Ae/Be stars
in nearby star forming regions. 
The best
studied young cluster in the solar neighborhood is the Orion Nebula Cluster
(ONC).
Its richness and high density make the ONC an
ideal target for the study of the stellar IMF 
from 45 M$_\odot$ to less than $\sim$0.02 M$_\odot$ (e.g., Hillenbrand
1997, Slesnick et al. 2004). 
The distribution of the optically visible stars in the HR diagram shown in
Fig.~\ref{onchrd} reveals several important features.  First, the upper
envelope of the stellar distribution is generally well matched by the
birthline over the full mass spectrum.  Second, stars more massive than
$\sim$8~M$_\odot$ are tightly grouped about the ZAMS, while less massive ones
diverge from it.  Third, the predicted endpoint of the birthline at M$\sim$8
M$_\odot$ also appears supported by the observational data:  there are no
massive stars still in the PMS in spite of the fact that many low-mass stars
have very young ages. Since there is evidence that the most massive stars are
also the youngest (see Sect.~5), the lack of massive PMS stars is quite
significant.

The ONC is not an exception and similar results are obtained in the Upper
Scorpius OB association, and in clusters such as NGC 6611, NGC 3603, and R136
in 30 Doradus. The HR diagrams displayed in the right panel of Fig.~\ref{hrd}
summarize the current knowledge on the intermediate-mass population in
several nearby ($\leq$500 pc) OB associations. Again, it is clear that most
stars are located between the birthline and the ZAMS, with the most massive
members already on the ZAMS. 

%

\section{PMS evolution of low-mass stars: lithium depletion and the age
spread of young clusters}

The initial evolution of stars with mass $\simle$2 M$_\odot$ is the least
affected by the protostellar conditions. The largest departure from the
standard theory of PMS stars concerns the history of deuterium burning since
the latter can effectively start already during the accretion phase for
masses above a few tenths of a solar mass.  
Thus, T Tauri
stars, having consumed most of the available deuterium during the accretion
phase, begin the contraction phase by simply descending along the vertical
portion of the Hayashi track.

More relevant is the next episode in the life of a PMS star, 
involving the nuclear burning and depletion of the second lightest isotope:
lithium (Li)
As contraction proceeds, the critical
temperature ($\sim 3\times 10^6$~K) for the reaction $^7$Li(p,$\alpha$)$^4$He
is reached, and the initial Li content is readily depleted in fully
convective, sub-solar stars (M$\lesssim$0.5 M$_\odot$).  Objects with mass
$\lesssim$0.065 M$_\odot$ never attain the ignition temperature and therefore
maintain their initial abundance.  The physics required to study the
depletion history as a function of age has little uncertainty, since it
depends mostly on the stellar mass and effective temperature 
(\cite{bil97}).  Detailed numerical models show that fully convective stars
in the range 0.5--0.2 M$_\odot$ start to deplete Li after about 5--10~Myr,
and completely destroy it in additional $\sim$10~Myr (e.g., \cite{bar98},
\cite{sie00}).  Lower mass stars take much longer to burn Li and, due to the
strong temperature sensitivity of the energy generation rate, there is a
sharp transition between fully depleted objects and those with the initial Li
content. Thus, Li-depletion can provide an independent clock to gauge stellar
ages.


%
\begin{figure}
  \centering
 \includegraphics[height=8cm,width=8cm,angle=0]{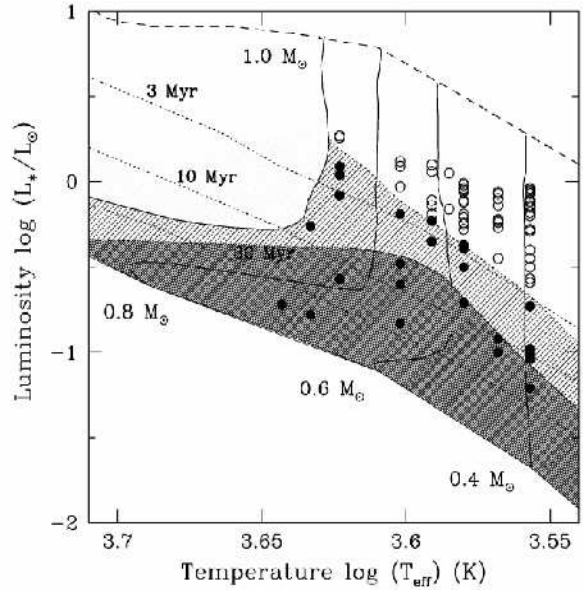}
  \caption{Distribution in the HR diagram of the 
low-mass ONC stars for which the lithium abundance has been measured.
The hatched regions indicate different levels of predicted Li depletion: up 
to a factor of 10 (light grey) and more (dark grey) below the initial value
 (using Siess et al. 2000 models).
 Selected masses and isochrones are indicated. Open and filled circles are
 for theoretically expected undepleted and depleted stars, respectively.
  }
  \label{lidepl}
\end{figure}
From the HR diagram of the ONC (Fig.~\ref{onchrd}), we see that the low-mass
population has a large spread in luminosity ($\Delta L\simgr$10). Does this
reflect a corresponding spread in stellar ages (factor of 10 or more), or
is it merely a scatter due to the observational uncertainties?
This is an important question since the presence of older stars in young
clusters bears directly on the issue of the duration of star formation.
It is still unclear
whether molecular clouds can sustain the production of
stars for an extended period of time ($t_{\rm cl}\approx 10^7$~yr), longer
than the typical free fall time ($t_{\rm ff}\approx 10^6$~yr) 
(e.g.~\cite{mac04}, \cite{tas04}). Therefore, the identification of an old
population in the ONC calls for an independent test, such as that provided
by Li-burning.

\begin{figure}
  \centering
 \includegraphics[height=7cm,width=7cm,angle=0]{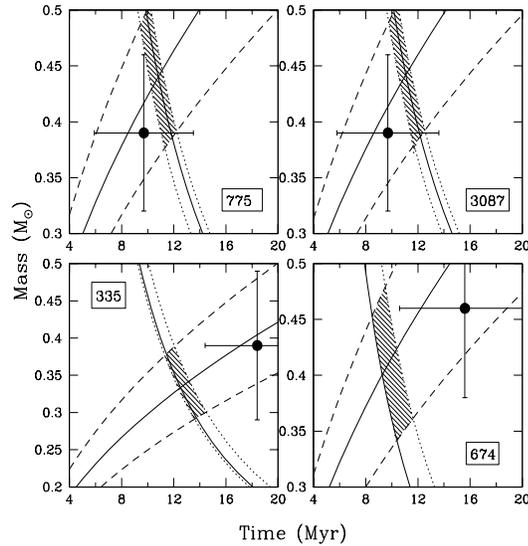}
  \caption{Mass vs. age diagram for the four stars with evidence for Li
 depletion. 
 In each panel, the solid points with
 errorbars give the mass and age from theoretical PMS tracks and isochrones.
 The hatched regions bound the values of mass and
 age consistent with the measured Li-abundances for the known 
 (L,T$_{\rm eff}$) values. (From \cite{pal05}) }
  \label{lidepl_bild}
\end{figure}

Figure~\ref{lidepl} shows the distribution in the HR diagram of a sample of
ONC members with mass in the range 0.4--1.0 M$_\odot$ and isochronal ages
greater than $\sim$1~Myr, together with the region of partial (light shading)
and full (dark shading) depletion predicted by stellar evolution models.
Lithium abundances have been derived for all the stars in Figure~\ref{lidepl}
from measurements of the Li~I~670.8 nm doublet observed with FLAMES on VLT
(\cite{pal05}). The main result is the discovery of significant depletion
(factor of 3 to 7) in four stars with estimated mass of $\sim$0.4~M$_\odot$
and age $\sim$10~Myr. Comparison with numerical and
analytical models shows excellent agreement (to within 10\%) between the
isochronal age and lithium depletion time scale for two objects, the first
case in lithium-poor PMS stars. This is shown in Figure~\ref{lidepl_bild}
where the values of mass and age obtained from the HR diagram (points with
error bars) are compared to the predictions of nuclear depletion following
the semi-analytic method of Bildsten et al.  The derived ages of about 10~Myr
indicate that the ONC does contain objects much older than the average age of
the dominant population.  Thus, the star formation history of the ONC extends
long in the past, although at a reduced rate, in accordance with the
empirical evidence found in the majority of nearby clusters and associations
(\cite{ps00}).  {\it Even the ONC did not come into existence in a single,
rapid burst.}

%
%

%
%


\section{PMS evolution of intermediate-mass stars: constraints from
pulsation and rotation}

Let us now consider the intermediate-mass stars in the range
2-10 M$_\odot$. Here, the impact of the protostellar initial consitions
on the early evolution is the most profound and needs to be carefully 
checked. In the following, I will discuss two tests that allow
to test the internal structure at the beginning of the PMS phase and 
the predictions of the protostellar mass-radius relation.

\subsection{Initial conditions from stellar pulsations}
\begin{figure}
  \centering
 \includegraphics[height=7cm,width=7cm,angle=0]{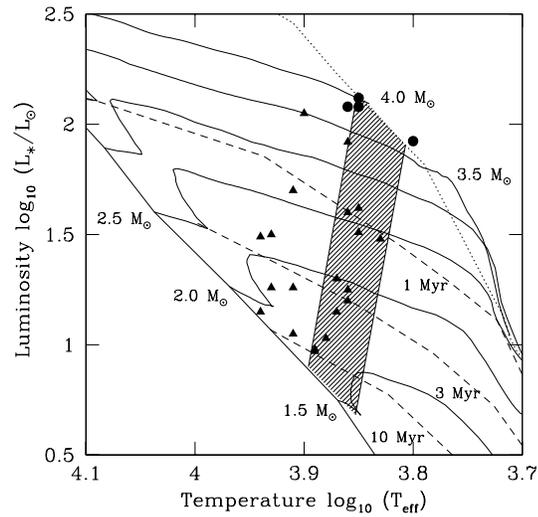}
  \caption{The position of the known Herbig Ae pulsators in the HR diagram. 
  The dashed region is the theoretical instability strip. The filled
  dots are the 4 stars that can be used to test the protostellar initial 
  conditions. (From \cite{mar04})
  }
  \label{stripcorot}
\end{figure}

On their way to the ZAMS, PMS stars with mass $1.5\simle~M/M_\odot~\simle 4$
cross the pulsation instability strip in the HR diagram. The instability is
related to the classical $\kappa$ and $\gamma$ mechanisms in the H- and
He-ionization regions near the stellar surface.  In spite of the relatively
short time spent in the instability strip (about 0.05--0.1 of the
Kelvin-Helmoltz time), more than 15 Herbig Ae stars have been identified as
$\delta$~Scuti-like pulsators with evidence for both linear and nonlinear
modes  (see the review by \cite{mar04}).  Their position in the HR
diagram is displayed in Fig.~\ref{stripcorot}, along with the instability
strip.  The typical pulsation periods are in the range of several hours and
amplitudes between hundredths and thousandths of magnitude.  These studies are
important since the relation between the pulsation period and the intrinsic
stellar parameters (L, T$_{\rm eff}$) allow to obtain independent information
on the evolutionary properties and in particular the stellar mass.
Additionally, at least in principle, asteroseismological techniques can be
applied to infer the internal structure of the observed pulsators.

\begin{figure}
  \centering
 \includegraphics[height=7cm,width=7cm,angle=0]{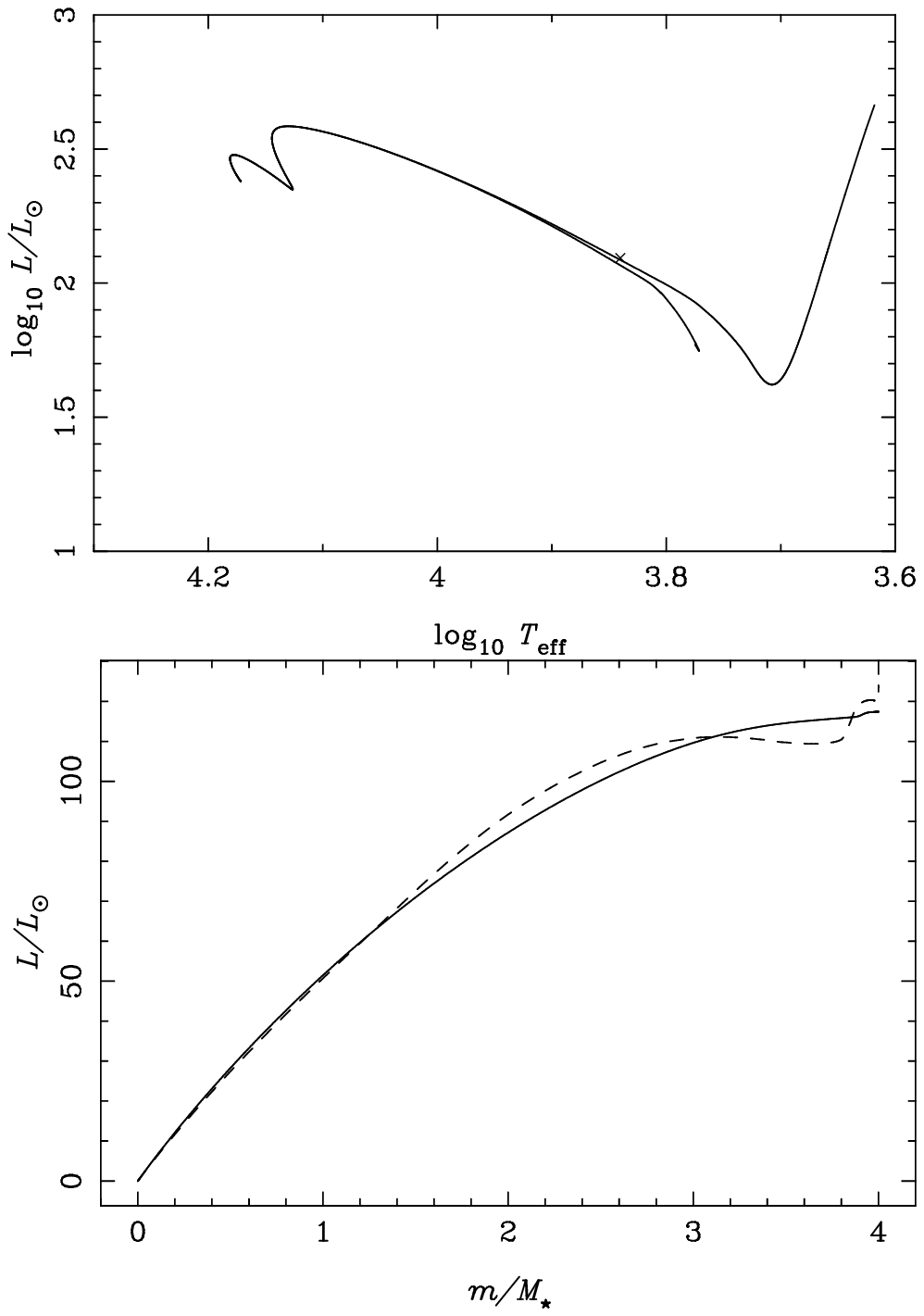}
  \caption{{\it Top}: Evolutionary tracks for a 4 M$_\odot$ star.
  The upper curve is for the standard initial conditions, while the
  lower curve is for the protostellar conditions. {\it Bottom}:
  The internal luminosity profile for the two models: the dashed curve is
  for the protostellar case. (Adapted from \cite{mar05}) }
  \label{figpuls}
\end{figure}

In the present context, it is of particular interest the group of 4 stars
with mass $\sim$4~M$_\odot$ located in the close vicinity of the birthline 
(see Fig.~\ref{stripcorot}). 
These stars represent an excellent sample to test
the initial conditions for PMS contraction. As shown in the upper panel of
Fig.~\ref{figpuls}, the track of a 4~M$_\odot$ star with the protostellar
initial conditions (lower curve) differs considerably from the standard one
(upper curve). The two curves join at a position near the red edge of the
instability strip. The lower panel of
Fig.~\ref{figpuls} shows the internal luminosity profile of the 4~M$_\odot$
star at the red edge: the solid and dashed lines are for the standard model
and protostellar initial conditions, respectively. The latter displays a
thermally unrelaxed luminosity profile while the former is characteristic of
a fully radiative object. Now, the pulsation characteristics (periods,
amplitudes) are particularly sensitive to the internal conditions, especially
in the subsurface regions where the perturbations are excited.  Detailed
modeling indicate large differences in the excited linear and nonlinear modes
(\cite{mar05}).  Hopefully, these predictions can be tested by high
sensitivity multisite monitoring campaigns and/or space observations (such as
the COROT satellite scheduled for launch in 2006).

\subsection{Effects of accretion on the rotational history}
The evolution of angular momentum in the PMS phase shows that rotation is an
important ingredient to understand young stars. Both T Tauri and Herbig Ae/Be
stars rotate at a small fraction ($\simeq$1/10) of the breakup velocity,
indicating that the angular momentum problem must be solved during the
protostellar phase (\cite{bou86}, \cite{fin85}). Yet, the
observed rotational velocities of PMS stars are larger than those of more evolved stars of similar
mass.  Thus, some angular momentum must be lost during PMS contraction.

Although wind braking is certainly present already in the PMS phase, it is
not expected to dominate the rotational evolution since it operates on a time
scale comparable or longer than $t_{\rm KH}$ (\cite{bou97}).
Presently, the most popular mechanism for the removal of angular momentum
involves magnetic star-disk interaction (so-called, disk-locking hypothesis;
\cite{kon91}; \cite{shu94}; see also \cite{mat04} for a
different view). According to this model, the stellar magnetic field threads
the circumstellar disk and truncates it at a characteristic radius
set by the balance between the pressure of the accreting
gas and the magnetic pressure in the disk.  Accretion of disk material onto
the star occurs along field lines, producing hot spots near the magnetic
poles and locking the stellar rotation to the angular (Keplerian) velocity of
the disk at corotation.  At the same time, magnetic torques transfer angular
momentum away from the star to the disk.  The location of the characteristic
radius varies with the accretion rate: if $\dot M_{\rm disk}$ increases, the
radius decreases, thus increasing the rotation speed of the
accreting gas and hence the star. Also, a higher accretion rate 
affects the mass-radius relation producing larger (proto)stellar radii.

Recently, \cite{str05} et al. have used this argument to propose an
interesting test on the initial conditions by
measuring the projected rotational velocities in early-type (B0--B9) stars in
the young clusters h and $\chi$ Persei and comparing them with those of a
sample of field stars of similar age ($\sim$15~Myr) and mass (4--15
M$_\odot$).  The basic idea is that if the accretion rates are higher in
dense clusters, the rotational velocities of the h the $\chi$ Persei cluster
should be higher on average than the field population born in loose
associations.  Indeed, Strom et al. find that the late B-type (4--9 M$_\odot$)
cluster stars rotate faster (factor of $\sim$2), while the difference is less
for the early-type B stars.  The higher rotation speed can be attributed to
the enhanced spinup experienced during contraction by stars formed with a
high accretion rate that produced large initial radii as a result of
deuterium shell burning. On the other hand, for more massive stars, the
difference in the initial radii are smaller due to their tendency to contract
while accreting.  If this evidence is corroborated by similar results in
other young clusters, it would provide a new way to test the predictions of
protostellar and PMS models.

\section{Early evolution of massive stars: clues from protostars}
What happens to massive stars? How can we test mass and age from the location
in the HR diagram? How can we validate the accretion scenario developed for
low- and intermediate-mass stars? Are very high accretion rates
($>10^{-4}$~M$_\odot$~yr$^{-1}$)) really required for massive star formation,
as repeatedly stated during the Conference?  And, in case, is there evidence
for an increase of $\dot M_{\rm acc}$ with mass? Since most massive stars are
in clusters, are they the first or the last to form?

Returning to the case of the ONC (see Fig.~\ref{onchrd}), we see that stars
with mass $\simgr$8 M$_\odot$ (the Trapezium stars) are aligned along
(actually, slightly to the right of) the ZAMS. Although all of them are part
of binary/multiple systems, none of them is an eclipsing binary. Thus, we
have no direct access to their mass and evolutionary status.  The only
exception is the least massive of the Trapezium stars, $\Theta^1$ Ori B (see
Fig.~\ref{onchrd}), whose components (each a binary system) have a dynamical
mass of M$_{p}$=6.3 M$_\odot$ and M$_{s}$=2.5 M$_\odot$, respectively. When
placed in the HR diagram, the secondary falls very close to the birthline and
the two stars align on the same isochrone for $t\sim 10^5$ yr, within the
observational uncertainties (\cite{ps01}). Thus, this somewhat massive system
is very young, confirming previous suggestions that massive stars tend to
form relatively late in a cluster. Unfortunately, current observations still
lack the required resolution to derive the stellar parameters of the
companions to the bright stars (although individual masses can be derived in
the next few years from orbit solutions). Clearly, future observations should
address this fundamental point.

Considering the formation mechanism, it is clear that the accretion scenario
that works well for lower mass objects finds some natural problems
at the highest masses. The main one is that radiation pressure begins to
become significant in protostars of intermediate mass. However, infall cannot
be stopped as long as the accretion component dominates the total
luminosity.  We have
seen that for $\dot M_{\rm acc}\sim 10^{-5}$~M$_\odot$~yr$^{-1}$, a
protostar joins the main sequence at a mass of $\sim$8 M$_\odot$ where it
releases $\sim$3000 L$_\odot$, less than the luminosity delivered by
H-burning ($\sim 10^4$~L$_\odot$) at that mass.  Hence, the need for higher
accretion rates (for example increasing with mass, e.g. \cite{ber01}),
indirectly suggested by the large mass {\it loss} rates inferred from the
associated outflows. How to achieve the physical conditions for such high
values of $\dot M_{\rm acc}$ is still not at all clear, although centrally
condensed turbulent cores may serve the purpose (\cite{mck03}, but see Dobbs
\& Bonnell, this volume). Perhaps, the early suggestion of an entirely
different mode of formation based on growth by coalescence of dense
molecular cores still represents a viable alternative (e.g.  \cite{sta00}). 
Interestingly, evidence for possible interactions between cores at high
infall rates has been found in the case of the dense clumps of NGC~2264C
(\cite{per05}).

We have reached an important conclusion:  in order to distinguish between the
two plausible explanations, massive protostars vs. young ZAMS stars, the
critical quantity to determine observationally is $\dot M_{\rm acc}$. Future
high angular resolution millimeter observations will reveal whether bright
embedded objects are characterized by {\it high} values of the accretion rate
(protostars) or not (ZAMS stars). Given the large distances of massive star
forming regions ($\simgr$1 kpc), one has to wait for the completion of
(sub)mm-wave arrays such as ALMA to reach the desired spatial resolutions.

\subsection{Unraveling a massive protostar: the case for Orion KL}
Owing to the extremely high extinction along the line of sight, the direct
observation of the source (photosphere or inner accretion disk) responsible
for the high infrared luminosity of the Orion BN/KL nebula is precluded.
However, the source can be indirectly seen through the scattered light of the
surrounding bipolar nebula (\cite{mor98}). Morino et al.  obtained a 2 $\mu$m
spectrum of the reflection nebula tracing the radiation that escaped the dust
surrounding the massive protostar possibly along the polar axis of a
circumstellar disk/torus. The spectrum shows absorption features due to CO
ro-vibration bands formed at T$\sim$4500~K.

If the observed 2~$\mu$m radiation is emitted by the protostellar
photosphere, the luminosity estimated from the reflected light and the
effective temperature yield a protostellar radius larger than $\sim$300
R$_\odot$. With such large radii, the required mass accretion rate onto the
protostar should exceed $\dot M_{\rm proto}\ge 5\times 10^{-3}$ M$_\odot$
yr$^{-1}$.  Alternatively, the inner portion of an accretion disk around a
massive protostar could be responsible for the observed near-infrared
radiation. For a steady accretion disk, the K-band radiation with an
effective temperature of 4500 K would be emitted by an annular region at a
distance of $\sim$200 R$_\odot$, assuming a central protostar of 10
M$_\odot$.  Given the size of the emitting region, accretion rates of $\dot
M_{\rm disk}\sim 10^{-2}$~M$_\odot$~yr$^{-1}$ are required in order to
produce to produce the observed K-band luminosity.

Both possibilities, a huge protostar or an extremely active disk, represent
extreme situations with respect to standard models of protostellar
evolution.  However, the difference between these pictures is large enough to
be tested observationally.  In particular, the CO bandhead {\it absorption}
lines at 1.6 and 2.3 $\mu$m are expected to have distinctively different
profiles in the two cases if observed at sufficiently high spectral
resolution ($\simgr$10000). This diagnostic differs from the study of CO {\it emission}
profiles observed towards a number of distant massive young stars (see
contribution by Blum) where the emission lines cannot be photospheric
in origin, but trace the circumstellar material close to the central object,
possibly in Keplerian rotation. In the case of the IR reflection nebula, the
{\it absorption} CO features could have an origin in the protostellar
photosphere and the velocity resolved profile can test this hypothesis on
this and similar objects.

\section{Conclusions}

The main points of this review are:

\begin{itemize}
\item Protostellar evolution via accretion  greatly affects the properties of
young PMS stars, particularly in the intermediate- and high-mass regime.

\item For low-mass PMS stars, Li-depletion during early PMS evolution
can be used to gauge the age and age spread in young clusters, 
such as the ONC.

\item The peculiar initial conditions of PMS stars with mass
$\sim$4--8~M$_\odot$ (Herbig Ae/Be) can be tested using pulsations and
rotation as reliable diagnostics.

\item The disk accretion scenario predicts that stars with mass
$\simgr$8--15~M$_\odot$ should be on the ZAMS once optically visible. 
Currently, there is no evidence for massive PMS stars. 

\item The empirical agreement of the position of the birthline in several
clusters/associations indicates that $\dot M_{\rm proto}\simle
10^{-5}-10^{-4}$~M$_\odot$~yr$^{-1}$ can account for the formation of stars
and their attendant disks up to $\sim$10--20~M$_\odot$.

\item For more massive protostars, the magnitude of $\dot M_{\rm proto}$ is
still unknown. CO bandhead {\it absorption} lines can 
be used to infer the accretion rate around massive protostars, the
most elusive quantity to measure. 

\end{itemize}



\end{document}